\documentclass[reprint,aps,prl,twocolumn]{revtex4-2}
\usepackage{amsfonts}
\usepackage[tbtags]{amsmath}
\usepackage{amssymb}
\usepackage{amsthm}
\usepackage{psfrag}
\usepackage{braket}
\usepackage[dvips]{epsfig}
\usepackage{epsfig}
\usepackage{subfigure}
\usepackage{color}
\usepackage{charter}
\usepackage{graphicx}
\usepackage{lipsum}
\usepackage{ulem}
\usepackage{hyperref}

\hypersetup{colorlinks=true, citecolor=blue, urlcolor=blue, linkcolor=black}

\begin{document}
\title{ Coherent Amplifier-Empowered Quantum Interferometer: Preserving
Sensitivity and Quantum Advantage under High Loss}
\author{Jie Zhao$^{1}$}
\author{Zeliang Wu$^{1}$}
\author{Haoran Liu$^{1}$}
\author{Yueya Liu$^{1}$}
\author{Xin Chen$^{1}$}
\author{Xinyun Liang$^{1}$}
\author{Wenfeng Huang$^{{1}}$}
\email{wfhuang@phy.ecnu.edu.cn}
\author{Chun-Hua Yuan$^{1}$}
\email{chyuan@phy.ecnu.edu.cn}
\author{L.Q.Chen$^{1,2,}$}
\email{lqchen@phy.ecnu.edu.cn}
\address{$^1$State Key Laboratory of Precision Spectroscop, Institute of Quantum Science and Precision Measurement, School of Physics, East China Normal University, Shanghai 200062, China}
\address{$^2$Shanghai Branch, Hefei National Laboratory,
	Shanghai 201315, China}

\begin{abstract}
Quantum interferometers offer phase measurement capabilities that surpass
the standard quantum limit (SQL), with phase sensitivity and quantum
enhancement factor serving as key performance metrics. However, practical
implementations face severe degradation of both metrics due to unavoidable
losses, representing the foremost challenge in advancing quantum
interferometry toward real-world applications. To address this challenge, we
propose a coherent-amplifier-empowered quantum interferometer. The coherent
amplifier dramatically suppresses the decay of both sensitivity and quantum
enhancement under high-loss conditions, maintaining phase sensitivity beyond
the original SQL even for losses exceeding 90\%. Using an injected 4.2 dB squeezed-vacuum state in experimental demonstration,  our scheme reduces the quantum enhancement degradation under 90\% loss from 3.7 dB in a conventional quantum interferometer (CQI) to only 1.5 dB. More importantly, the phase sensitivity degradation under the same loss is limited to 4.0 dB, markedly outperforming the 11.2 dB degradation observed in a CQI. This improvement is enabled by the coherent amplifier’s phase-sensitive photon amplification and its protection of the quantum state. 
This breakthrough in amplifier-empowered quantum interferometry overcomes the critical barrier to practical deployment, enabling robust quantum-enhanced measurements in lossy environments.
\end{abstract}

\maketitle

\preprint{APS/123-QED}

Quantum interferometry is a high-sensitivity measurement technology and has
been widely applied to the measurement of various physical parameters, such
as atomic gravimeters {\cite{gravity1,gravity2}} and atomic gyroscopes {\cite%
{atomgyroscope1,atomgyroscope2, atomgyroscope3,atomgyroscope4}} based on
atomic interferometry {\cite{a}}, gravitational wave measurements {\cite%
{LIGO,LIGO1,LIGO2,LIGO3}} and optical gyroscopes {\cite%
{optgyroscope1,optgyroscope2}} based on optical interferometry. Phase
sensitivity $\delta \phi $, the core indicator of interferometer
performance, is always limited by the standard quantum limit (SQL) {\cite%
{SQL, SQL1, SQL2}} $\delta \phi _{SQL}=1/\sqrt{N}$ for coherent
interferometers and can be enhanced to $\delta \phi =1/(M\sqrt{N})$ for
quantum one, where $N$ is the phase-sensitive particle number and $M$ is the
quantum enhancement factor due to quantum states {\cite{Caves2, J.P.Dowling}}%
. Therefore, quantum interferometers (QI) have the advantage of phase sensitivity
exceeding SQL {\cite{S1, SU11, Truncated, SU112, NOON1, NOON2, NOON3, S2, S3}%
}. To enhance sensitivity, larger photon numbers $N$ and higher-quality
quantum states with larger $M$ are essential. Over the past decade, many
schemes have been proposed and demonstrated to improve the performance of
quantum interferometers. These include enhancing the quantum properties of
the states \cite{Boto}, establishing quantum correlations between the
two interferometer arms {\cite{SU11,Truncated,SU112}}, integrating
 parametric amplifiers into each arm of the laser interferometer to generate quantum squeezed states {\cite{Jia}}, and implementing SU(1,1)-SU(2) nest interferometry {\cite{Du}}. Most of these studies have focused on the improvement of quantum enhancement under
lossless conditions.

However, in practical applications, unavoidable losses quickly reduce $N$
and the quantum enhancement factor $M$, resulting in quantum interferometers rapidly
losing their quantum advantage and obtaining a bad phase sensitivity \cite{Demkowicz-Dobrzanski, Ruchbah, Xin, Zhang}. For the CQI shown in Fig. 1(a), when a 10 dB-squeezed vacuum field is injected, the quantum enhancement factor $M$ can
reach 10 dB under the lossless condition, corresponding to a 10 dB improvement
in phase sensitivity. However, when the interference arm experiences a large
loss of 90$\%$, the $M$ factor degrades 9.2 dB, while $N$ exhibits a
significant reduction. This results in a total phase sensitivity degradation
of 16.6 dB, which is far worse than the SQL. The large loss poses a
significant challenge for the practical application of quantum
interferometry. The large loss poses a significant challenge for the practical application of quantum interferometry.  
Specifically, the resulting degradation in sensitivity and quantum enhancement fundamentally limits the applicability of quantum interferometers to critical areas such as satellite-based quantum links \cite{liao2017long,liao2017satellite}, resource-constrained sensing networks \cite{guo2020distributed,malia2022distributed}, and photon-damage-sensitive biological imaging \cite{topfer2022quantum,jacques2013optical}.

To address optical loss impact on quantum enhancement, Huang et al. {\cite%
{Huang}} proposed a variable beamsplitter to protect quantum enhancement
factor $M$, while Frascella G. et al. \cite{Frascella} utilized quantum
amplifiers to improve $M$ under high loss. Although both methods mitigate $M$%
-factor degradation, their phase sensitivities exhibit significant declines.
So far, no experimental demonstration has achieved concurrent preservation
of quantum-enhanced $M$ and phase sensitivity (corresponding to $N$) under
large-loss conditions, posing a fundamental barrier to quantum
interferometers in practical sensing applications.

\begin{figure}[tbph]
\begin{centering}
		\includegraphics[width=0.84\columnwidth]{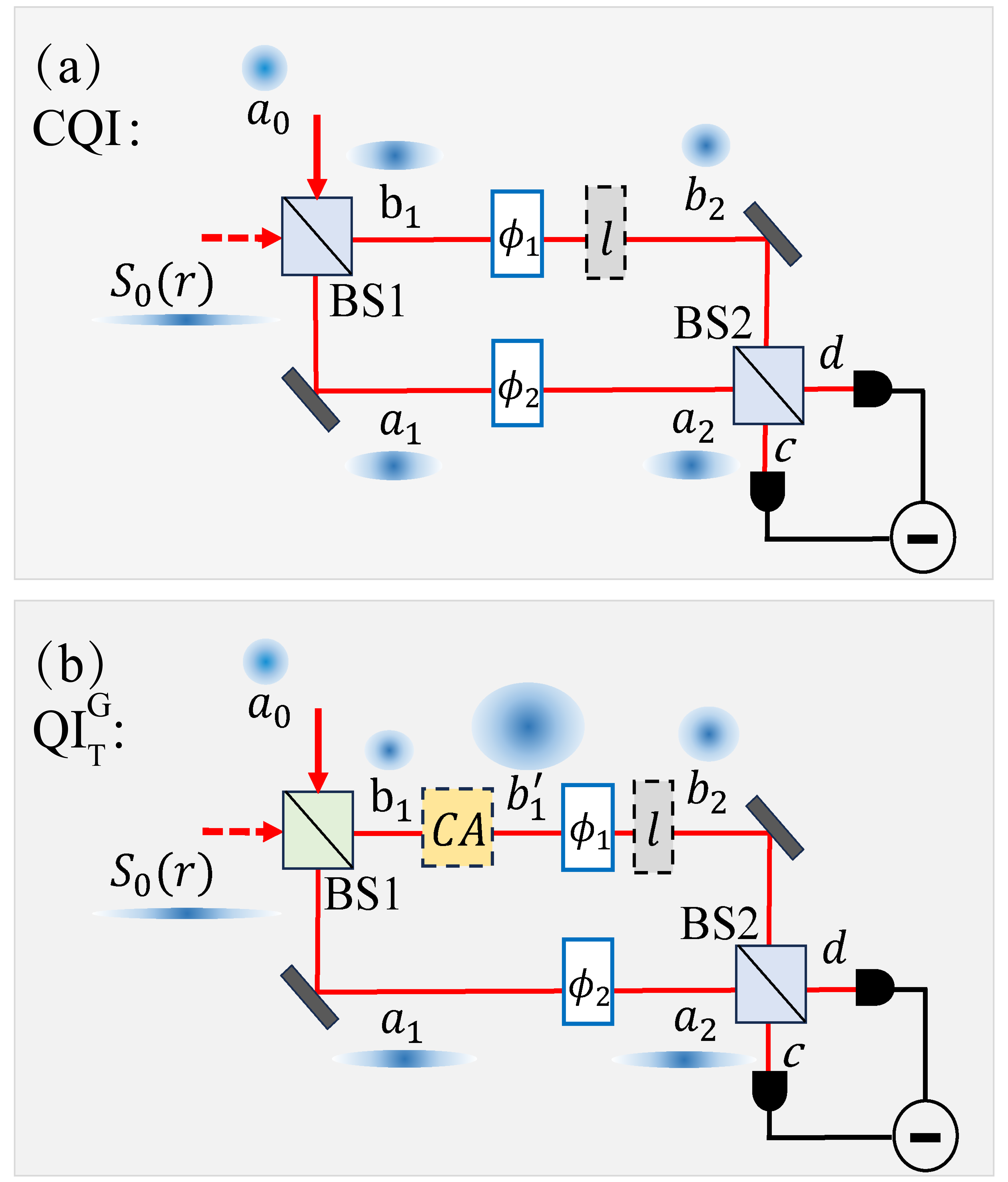} 
		\par\end{centering}
\caption{Quantum interferometer schemes. (a) CQI: conventional quantum
interferometer. The splitting ratio of the two beam splitters, BS1 and BS2,
is 0.5:0.5, respectively. (b) $\mathrm{{QI_{T}^{G}}}$: quantum interferometer
integrated with a coherent amplifier (CA) of gain factor $G$ and BS1 of
adjustable splitting ratio $T:(1-T)$. The splitting ratio of BS2 is 0.5:0.5. $a_{0}$%
: input coherent state; $S_{0}(r)$: input squeezed vacuum state; $r$:
squeezing degree; $l$: optical loss rate of arm $b_{1}$; $\protect\phi=%
\protect\phi_{1}-\protect\phi_{2}$: phase shift. Blue shades: phase-space noise at each
position.}
\end{figure}

In this paper, to against the impact of large loss on phase sensitivity and
quantum enhancement, we propose and demonstrate a quantum interferometer
empowered by a coherent amplifier (CA) with gain factor $G$ and a
beam splitter with adjustable splitting ratio, signed as QI$_{T}^{G}$.
The diagram of QI$_{T}^{G}$ is shown in Fig. 1(b). A coherent field ($a_{0}$)
and a squeezed vacuum field ($S_{0}$) are combined on the beam splitter BS1 whose transmission ratio is $T$, forming two interference arms $a_{1}$ (reference
arm), $b_{1}$ (signal arm). Then $b_{1}$ passes the CA, couples a
phase shift $\phi _{1}$ and experiences optical loss $l$. Finally, two arms
recombine and interfere at the beam splitter BS2 whose splitting ratio is
0.5:0.5. The main difference between CQI and QI$_{T}^{G}$ lies in the
adoption of CA, and the adjustable splitting ratio of BS1. The CA plays a pivotal role in
counteracting loss-induced degradation, which dramatically suppresses the
decay of the quantum enhancement and phase sensitivity as loss increases.
Moreover, through optimizing the splitting ratio of BS1, it can
further protect quantum resources. The integration of CA and BS1 ensures
robust performance of the quantum interferometer even in high-loss
scenarios. Compared with CQI, under 90$\%$ loss, QI$_{T}^{G}$ can significantly reduce phase sensitivity degradation from 16.6 dB to 6.7 dB, and resist the decline of the $M$ factor from 9.2 dB to 5.0 dB. Below, we will theoretically
analyze and experimentally demonstrate it in detail.

In theory, the interference signal and noise of QI$_{T}^{G}$ can be written
as (Supplementary Material, Sec. I \cite{supplementary}): 
\begin{flalign}
Signal &=2\sqrt{(1-l)(1-T)T}GN, \\
Noise &=\sqrt{[G^{2}(1-l)(T+e^{-2r})+T(2l-1)]N},
\end{flalign}%
which depends on the gain factor $G$, optical loss rate $l$, squeezing
degree $r$, the transmission ratio $T$ of BS1 and the photon number $N$ of the
input laser $a_{0}$. In actual scenarios, the number of input photons $N$
and the squeezing degree $r$ are always finite \cite{LIGO1, LIGO2, LIGO3}.
Therefore, we can optimize parameters $G$ and $T$ to improve the performance
of the quantum interferometer in a loss environment as much as possible. The
optimized $G$, $T$ values, $G_{opt}$ and $T_{opt}$, for optimal sensitivity $%
\delta \phi _{opt}$ are given as:

(i) When $0\leq l\leq 0.5$, $G_{opt}=1$, 
\begin{flalign}
	T_{opt} &= 
	\begin{cases} 
	\varsigma+\sqrt{\varsigma^{2}-\varsigma}&, (0< l\leq 0.5, \varsigma=(1-\frac{1}{l})e^{-2r}) \\
	0.5&,  (l=0).\\ 
	\end{cases}\\
\delta \phi _{opt}&=\frac{\sqrt{T_{opt}l+(1-l)e^{-2r}}}{\sqrt{%
4(1-T_{opt})T_{opt}(1-l)}}\cdot \frac{1}{\sqrt{N}}.&
\end{flalign}

(ii) When $0.5<l<1$, $G_{opt}\rightarrow +\infty $, $T_{opt}=e^{-2r}(\sqrt{%
1+e^{2r}}-1)$, and 
\begin{flalign}
\delta \phi _{opt}&=\sqrt{\frac{e^{-2r}+T_{opt}}{4(1-T_{opt})T_{opt}}}\cdot 
\frac{1}{\sqrt{N}}.&
\end{flalign}%
There is a transition point between the two cases, determined by the
splitting ratio of the BS2 (Supplementary Materials, Sect. II \cite%
{supplementary}). For the QI$_{T}^{G}$ protocol implementation with a 0.5:0.5
BS2, this critical transition occurs at $l$=0.5. $G_{opt}$\ is entirely
different in the two cases, arising from the competition between signal
amplification and noise growth.\textbf{\ } In case of $0\leq l\leq 0.5$, $G_{opt} =1$ implies that no amplifier is required, mainly due to the noise outpacing signal magnification. When $0.5<l<1$, $G_{opt}\rightarrow +\infty$ means that the larger $G$ is, the better the sensitivity will be. The signal
magnification outpaces the noise growth. This can be clearly seen in Fig. 2.
The total noise includes amplification-induced noise (noise-$G$) as well as
loss-induced vacuum noise (noise-$l$). Fig. 2(a, b) shows that the signal
(signal-$G$) exhibits continuous growth with the gain $G$. At a low loss rate ($l$=0.2) in Fig. 2(a), noise-$G$ dominates over noise-$l$, causing the total noise to amplify faster than the signal. However, when the loss rate exceeds 50$\%$ ($l$=0.9) in Fig. 2(b), noise-$l$ rises progressively and becomes comparable to even larger than noise-$G$. In this regime ($l>0.5$), the growth rate of the signal-$G$ surpasses the total noise increase rate. The specific expressions can be seen in the Supplementary Material, Sec. III \cite{supplementary}. 

\begin{figure}[tbph]
\begin{centering}
		\includegraphics[width=1.0\columnwidth]{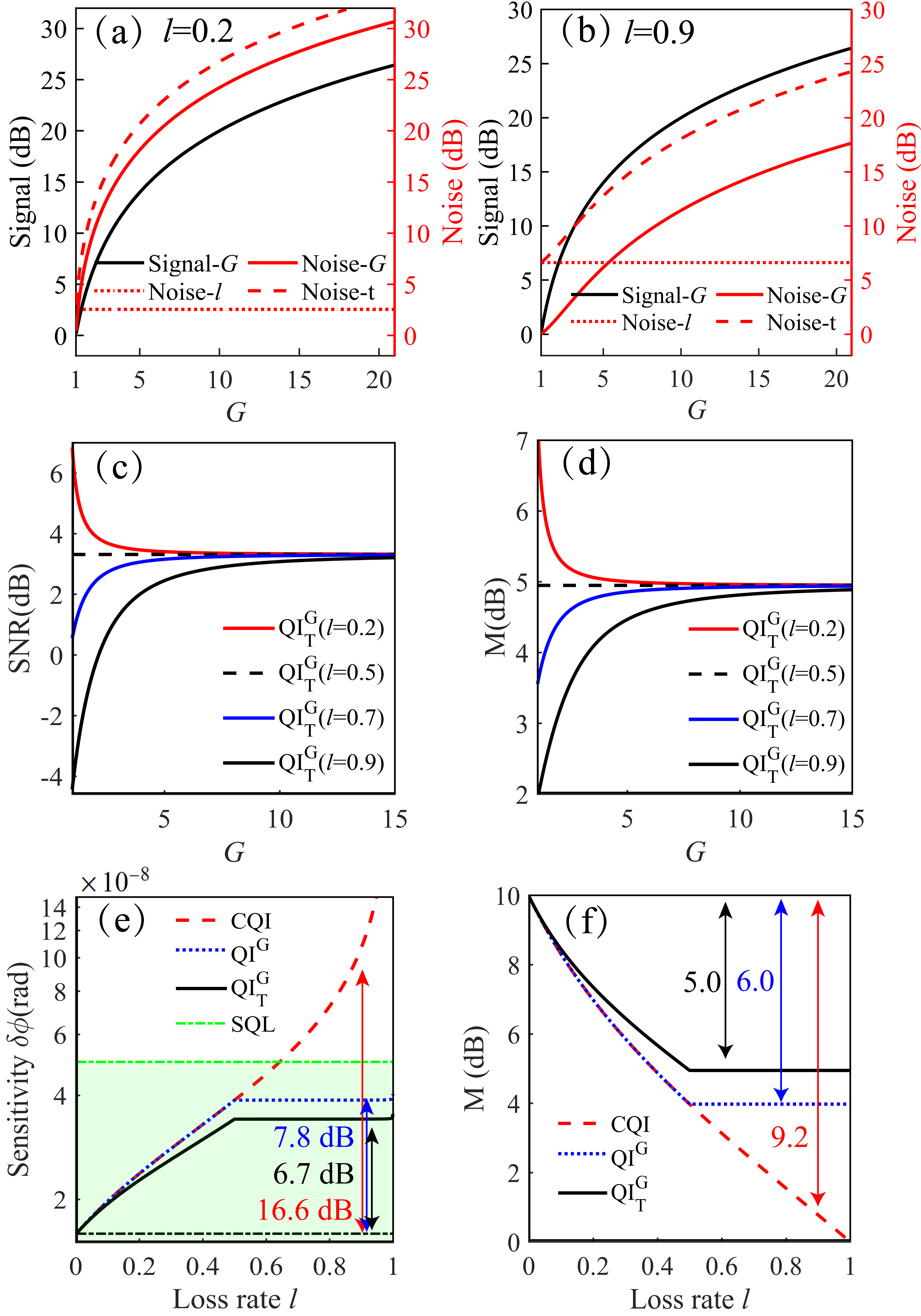} 
		\par\end{centering}
\caption{Theoretical results. (a-b) Signal and noise versus gain $G$ for QI$%
_{T}^{G}$ at $l$ = 0.2 and 0.9, respectively. Signal-$G$: the amplified
signal; noise-$G$: the amplification-induced noise; noise-$l$: the
loss-induced noise; noise-$t$: total noise, the sum of noise-$G$ and noise-$l
$. (c-d) The relative signal-to-noise ratio (SNR) and quantum enhancement $M$
as functions of gain $G$ in QI$_{T}^{G}$, plotted for different loss rates $l
$. The zero level is the result of CQI at $r=0, l=0$. (e-f) Optimal
sensitivity and quantum enhancement versus loss rate $l$. Dashed, dotted and
solid curves correspond to CQI, QI$^{G}$ ($G_{opt}$ and T=0.5), and QI$%
_{T}^{G}$ ($G_{opt}$ and $T_{opt}$), respectively. The green and black dash-dotted lines in (e) represent the SQL and the lossless CQI sensitivity, respectively. The SQL is obtained by $1/\protect\sqrt{N}$ with $N=4\times 10^{14}$. The squeezing degree of the injected squeezed-vacuum state is 10 dB (r=1.15). $M=-20\log _{10}[\protect\delta \protect\phi _{r}/\protect\delta \protect\phi _{r\rightarrow
0}]$.}
\end{figure}

In Fig. 2 (c,d), we identify a fundamental transition in amplifier behavior
at $l$=0.5, where signal and noise amplification exactly balance, leading to
both signal-to-noise ratio (SNR) and quantum enhancement factor $M$ become
independent of the gain factor $G$. Below this threshold ($l<0.5$),
increasing the gain $G$ degrades performances ($\delta \phi $ and $M$)
because the noise dominates, driving the system approaching the transition
point while reducing phase sensitivity and quantum enhancement factor. Above
threshold ($l>0.5$), the signal-to-noise ratio (SNR) and quantum enhancement
($M$) initially show poor performance. However, with increasing gain $G$,
signal amplification becomes dominant, significantly enhancing both SNR and $M$ until they approached critical-point values. This sharp transition reveals that amplifiers, while harm low-loss operation, are indispensable for
high-loss quantum interferometry, with performances fundamentally bounded by
the $l$=0.5 critical point. Fig. 2 (e,f) clearly shows that while CQI
suffers rapid performance decay with increasing loss, QI$_{T}^{G}$ maintains
the performance at the 50$\%$ loss level. Most significantly, when the loss
rate exceeds this critical threshold 50$\%$, the coherent amplifier exhibits
the capability to completely suppress any further degradation of $\delta \phi $ and $M$ regardless of additional losses.

The optimal sensitivity $\delta \phi _{opt}$ and the quantum enhancement factor $M$ of QI$%
_{T}^{G}$ are theoretically analyzed as functions of loss rate $l$, with
comparative results for CQI and QI$^{G}$ shown in Fig. 2(e-f),
respectively. In all cases, both phase sensitivity and quantum enhancement
degrade as the loss rate increases. Notably, CQI exhibits the most rapid
deterioration. For instance, under conditions with a 10 dB squeezed state
and 90$\%$ loss in the interference arm, CQI's quantum enhancement $M$ drops
from 10 dB to 0.8 dB, while its phase sensitivity degrades by 16.6 dB. The
utilization of CA (as seen in the QI$^{G}$ curves) reveals a distinct
behavior: a sharp transition occurs at $l$=0.5. Below this threshold, both
phase sensitivity ($\delta \phi $) and quantum enhancement factor ($M$)
exhibit rapid deterioration, whereas above the critical point they maintain
remarkable stability with negligible degradation. Therefore, $G_{opt}$ plays
a critical role in both improving phase sensitivity and preserving quantum
states. Specifically, under $l=90\%$ condition, compared with the CQI scheme, the QI$^{G}$ scheme has improved the phase sensitivity by 8.8 dB and increased the M factor by 3.2 dB.  Furthermore, optimizing the transmission ratio $T$ of BS1 provides only marginal gains in phase sensitivity but effectively preserves quantum states, yielding additional improvements of 1.1 dB for $\delta \phi$ and 1.0 dB for $M$. Overall,
compared to the CQI case, QI$_{T}^{G}$ achieves a total sensitivity enhancement
of 9.9 dB and a 4.2 dB improvement in the quantum enhancement factor $M$. The detailed data are provided in Tab. I of the Supplementary Material, Section IV \cite{supplementary}. Moreover, the amplifier's degradation-mitigation advantage becomes increasingly pronounced at higher loss levels. When the loss rate attains as high as 99$\%$, the QI$_{G}^{T}$ scheme achieves a 4.9 dB enhancement in the $M$ factor and a 20.3 dB improvement in sensitivity. Remarkably, the phase sensitivity of QI$_{G}^{T}$ maintains a 3.3 dB surpassing the original SQL even at near-total loss ($l\approx99\%$) in Fig. 2(e), primarily due to the benefits of $G_{opt}$. This coherent amplifier
implementation effectively overcomes the long-standing loss limitation in
quantum interferometry. We experimentally validate these advantages in the
following section.

\begin{figure*}[tbph]
\begin{centering}
		\includegraphics[width=1.75\columnwidth]{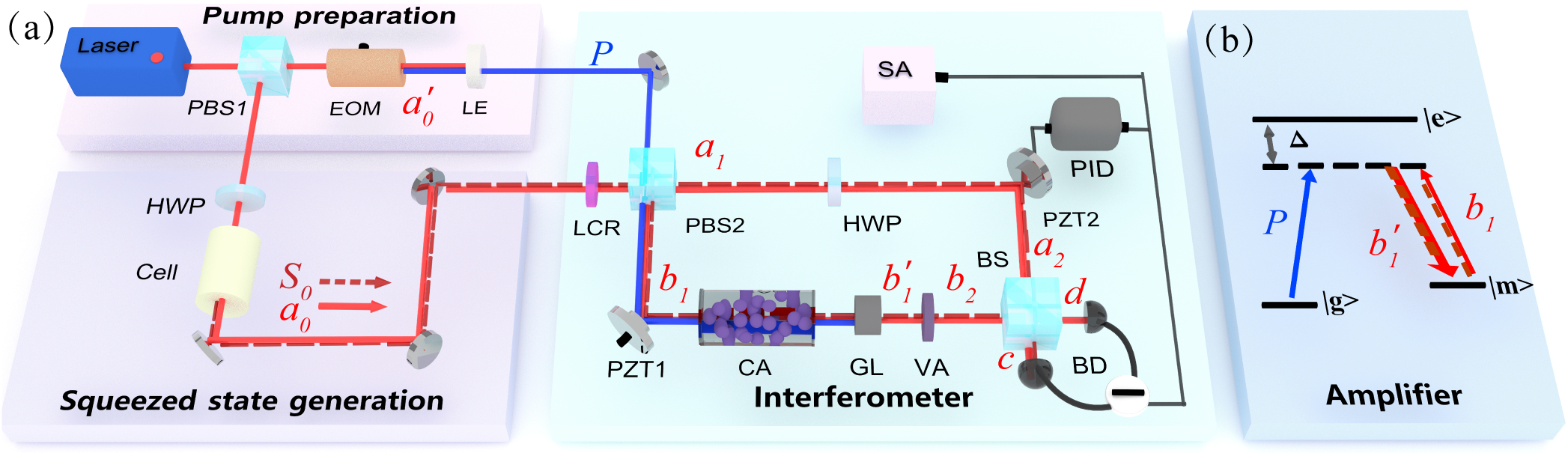}
		\par\end{centering}
\caption{Experimental setup of QI$_{T}^{G}$. (a) Experimental setup. Squeezed state generation is that the coherent field $a_{0}$ (red solid line, horizontal polarization) generates the squeezed vacuum field $S_{0}$ (red dashed line, vertical polarization) in the $^{87}$Rb atomic cell via polarization self-rotation effect. The frequencies and spatial modes of the $a_{0}$ and $S_{0}$ fields are the same. HWP: half-wave plate; PBS: polarization beam splitter; BS: beam splitter; EOM: electro-optic modulator; LE: laser etalon (filter out the unnecessary light); PZT: piezoelectric transducer; CA: coherent amplifier; GL: Glan polarizer (filtering the excessive pump field); VA: variable attenuator, simulating the loss $l$; M: mirror; PD: photoelectric detector; BD: balanced detection of intensity-difference; PID: proportional-integral-derivative phase-locking device; SA: spectrum analyzer;  $P$: the pump field (blue line, horizontal polarization) for CA. This $P$ field derives from the $a_{0}^{\prime }$ after a 6.83 GHz frequency shift via the EOM.  (b) Atomic energies of the CA. $b_{1}^{\prime }$ is the amplified $b_{1}$. $\left\vert g\right\rangle $: $\left\vert 5^{2}S_{1/2},F=1\right\rangle $; $\left\vert m\right\rangle $: $\left\vert
5^{2}S_{1/2},F^{\prime }=2\right\rangle $; $\left\vert e\right\rangle $: $%
\left\vert 5^{2}P_{1/2},F^{\prime \prime }=1\right\rangle $. $\Delta =600$
MHz.}
\end{figure*}

Fig. 3 presents the experimental optical configuration for implementing QI$%
_{T}^{G}$. The system begins with a laser beam divided into two components ($%
a_{0}$ and $a_{0}^{\prime }$). The $a_{0}$ beam undergoes squeezed state
generation (SSG) through polarization self-rotation in an atomic vapor cell 
\cite{Barreiro,Zhao}, producing a 4.2 dB squeezed vacuum field $%
S_{0}(r)$. The co-propagating $%
a_{0}$ ($100 \mu\rm{W}$) and $S_{0}$ fields, having the same frequency but orthogonal
polarization states, are directed to the beam splitter unit\textbf{\ }BS1. This optical assembly consists of a liquid crystal retarder (LCR) and a
polarization beam splitter (PBS2), with its transmission ratio $T$ precisely
controlled via LCR. The output forms two interference arms: $a_{1}$
(reference arm) and $b_{1}$ (signal arm). The signal arm ($b_{1}$) first
passes through a coherent amplifier (CA) that produces amplified output $%
b_{1}^{\prime }$, followed by a variable optical attenuator (VOA) with
adjustable loss $l$, yielding the attenuated beam $b_{2}$. This beam then recombines with the reference arm $a_{2}$ at BS2, creating interference fields ($c$ and $d$) that are detected by a balanced detector
(BD). The CA utilizes stimulated Raman amplification \cite{raymer1981stimulated, PhysRevLett.115.043602,hammerer2010quantum} in an $^{87}$Rb atomic vapor, with the gain factor $G$ precisely controlled through the intensity
of the Raman pump field ($P$). The relative phase difference between the two interference arms is locked at $\frac{\pi }{2}$ using a piezoelectric transducer
(PZT2) with servo bandwidth of DC-1 kHz. As the signal to be measured, a tiny phase-shift of $5\times10^{-4}$ rad (at 1 MHz) is introduced through PZT1. The resolution bandwidth (RBW) and video bandwidth (VBW) of SA are 30 kHz and 30 Hz, respectively. The quantum enhancement factor $M$ is quantified by comparing the interference noise levels between two measurement configurations: with and without an injected squeezed field.

The signal, noise, phase sensitivity $\delta \phi $ and quantum enhancement
factor $M$ are measured as a function of loss rate $l$ for three cases (CQI,
QI$^{G}$ and QI$_{T}^{G}$) with results given in Fig. 4(a-d). In case of
CQI, the signal quickly decreases and the noise increases with the loss rate 
$l$. As a result, under injection of a 4.2 dB squeezed vacuum field, the
phase sensitivity decreases rapidly with increasing loss, transitioning from
4.2 dB beyond the SQL ($l=0$) to a performance of nearly 7.0 dB below SQL ($l=0.9$%
). The quantum enhancement $M$ drops from 4.2 dB ($l=0$) to 0.5 dB ($l=0.9$%
). The experimental result aligns well with the theoretical prediction.

\begin{figure}[thbp!]
\begin{centering}
	\includegraphics[width=1.01\columnwidth]{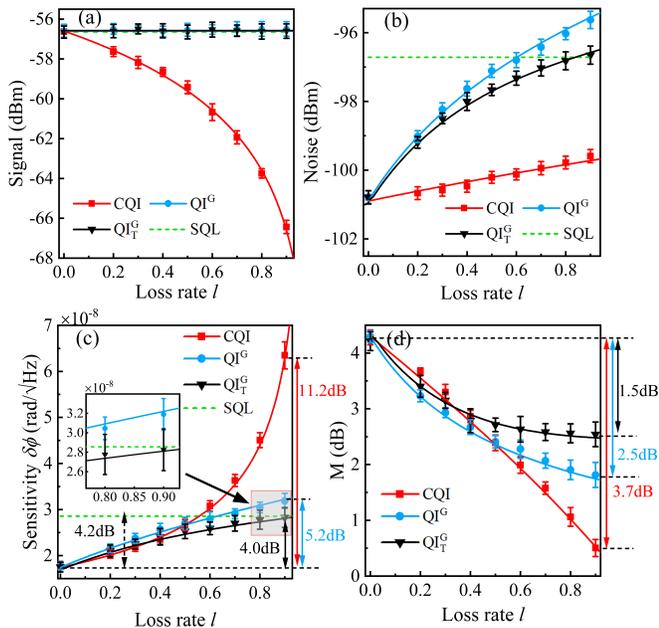}
	\par\end{centering}
\caption{Experimental results. The signal (a), noise (b), phase sensitivity
(c) and quantum enhancement (d) as functions of loss rate $l$ for CQI (red
square), QI$^{G}$ (with $G_{opt}$, $T=0.5$, blue dot), and QI$_{T}^{G}$ (with $G_{opt}$ and $T_{opt}$, black triangle). The squeezing degree of the input squeezed-vacuum field is 4.2 dB. The solid lines represent the theoretical results with parameters $N=1.2\times 10^{15}, r=0.48$ (i.e., 4.2 dB squeezing).}
\end{figure}

Theoretical analysis predicts that employing a coherent amplifier with
optimized gain $G_{opt}\rightarrow \infty $ under high-loss conditions would
yield substantial performance improvements. However, practical limitations
prevent the realization of such an infinite gain $G$. To enable meaningful
comparison and demonstrate the advancements of current QI$_{T}^{G}$, we
adopt the same phase-sensitive photon number condition in this experiment,
that is, the photon number entering the BD remains the same ($G, T$ satisfies: $2G\sqrt{(1-l)(1-T)T}=1$). Under this
condition, the signal of QI$_{T}^{G}$ remains the same as the lossless one
(the dotted curve in Fig. 4(a)) despite increasing loss, whereas its noise
gradually accumulates, consistently exceeding that of CQI due to
amplification and loss effects (the dotted curve in Fig. 4(b)). The coherent
amplifier demonstrates remarkable mitigation of degradation in phase
sensitivity $\delta \phi$ and quantum enhancement $M$, as evidenced in Fig.
4(c-d) - a direct consequence of signal amplification outweighing noise
growth. As a result, at $l=0.9$, even only optimizing the gain G, QI$^{G}$ demonstrates 6.0 dB better sensitivity and 1.2 dB greater quantum enhancement M compared to the CQI case.

Furthermore, optimizing the beam-splitter ratio enables enhanced quantum
state protection. As shown in Fig. 4(c-d), the quantum enhancement factor
increases by about 1.0 dB via $T_{opt}$ at $l$=0.9, yielding an
approximately 1.2 dB enhancement in sensitivity. The detailed data are provided in Tab. II of the Supplementary Material, Section IV \cite{supplementary}. While the optimized splitting ratio helps retain more quantum resources, it provides minimal compensation for the absolute phase sensitivity degradation caused by
optical loss---consistent with theoretical predictions.

Ultimately, compared with CQI, QI$_{T}^{G}$ achieves a total phase sensitivity improvement of 7.2 dB while maintaining a 2.7 dB quantum enhancement at 90$\%$ loss. This 7.2 dB sensitivity optimization is nearly equivalent to that
of a Mach-Zehnder interferometer (MZI) operating with 5.2 times input coherent photon number $N$. The QI$_{T}^{G}$ has exceptional performance in maintaining phase sensitivity and protecting quantum enhancement in the high-loss environment, as well as to maximize the utilization of quantum squeezing resources required for a QI.

In this paper, we present and experimentally validate a quantum
interferometer integrated with a coherent amplifier and a beamsplitter with
adjustable splitting ratio, addressing a fundamental challenge in quantum
interferometry: the rapid degradation of phase sensitivity $\delta \phi$
and quantum enhancement $M$ in the lossy environment. This degradation
represents a critical barrier to transitioning quantum measurement technologies from laboratory demonstrations to real-world applications. Our  results demonstrate that the coherent amplifier dramatically suppresses the
decay of both sensitivity and quantum enhancement under high-loss
conditions. Theoretical analysis shows that with a 10 dB squeezed vacuum
field and fixed photon number $N$, the system maintains $M=5.0$ dB and
sensitivity beyond the original SQL 3.3 dB even under 99$\%$ optical loss.
Experimental verification using a 4.2 dB squeezed field demonstrates the effectiveness of the QI$_{T}^{G}$ scheme, sustaining a beyond-SQL sensitivity and a 2.7 dB quantum enhancement $M$ at $l=90\%$. Compared to CQI, our coherent amplifier offers superior loss tolerance with a significantly simpler implementation. This quantum interferometry optimization design has successfully overcome the loss impacts in a high-loss environment, and it is expected to overcome the barrier preventing the transition of quantum interferometer technology from the laboratory to practical applications.

\section*{Acknowledgments}

This work is supported by the National Natural Science Foundation of China
Grants No. U23A2075, No. 12274132, No. 12104161, No. 12574391, and No. 12504564; the Natural Science Foundation of Shanghai Grant No. 25ZR1401112; Quantum Science and Technology-National Science and Technology Major Project grants 2021ZD0303200; Fundamental Research Funds for the Central Universities, and the Innovation Program of Shanghai Municipal Education Commission No. 202101070008E00099; the China Postdoctoral Science Foundation Grants No. 2023M741187, No. GZC20230815, and No. 2025T180920; the Shanghai Science and Technology Innovation Project Grant No. 24LZ1400600.

\bibliographystyle{apsrev4-2}
\bibliography{apssamp}

\end{document}